\documentclass{PoS}

\usepackage{graphicx}
\usepackage{amsmath}
\usepackage{amssymb}
\usepackage{media9}

\usepackage[caption=false]{subfig}

\usepackage{url}

\let\OLDthebibliography\thebibliography
\renewcommand\thebibliography[1]{
  \OLDthebibliography{#1}
  \setlength{\parskip}{0pt}
  \setlength{\itemsep}{3pt plus 0.3ex}
}

\title{Publicising Lattice Field Theory through Visualisation}

\ShortTitle{Lattice Field Theory Visualisation}
\author{James Biddle, 
Josh Charvetto,
Waseem Kamleh,
\speaker{Derek Leinweber},
Helen Piercy,
Ethan Puckridge,
Finn Stokes,
Ross D. Young${}^*$,
James Zanotti\thanks{This research is supported with supercomputing resources
    provided by the Phoenix HPC service at the University of Adelaide and the National
    Computational Infrastructure (NCI) supported by the Australian Government. This research is
    supported by the Australian Research Council through Grants No.\ DP140103067, DP150103164, and
    LE160100051.} 
\\
Centre for the Subatomic Structure of Matter, Department of Physics,\\ The University of Adelaide, SA 5005, Australia\\
E-mail: \email{derek.leinweber@adelaide.edu.au}}

\abstract{ 
The gluon field configurations that form the foundation of every
lattice QCD calculation contain a rich diversity of emergent
nonperturbative structure.  Visualisations of these phenomena not only
serve to explain the concept of a nontrivial vacuum but also entertain
a diverse audience from research funding panels to the next generation
of science enthusiasts.  In this brief review, a collection of
QCD-vacuum visualisations is presented including the structure of
chromo-electromagnetic fields, centre-cluster evolution at finite
temperature, the structure of projected centre vortices, and novel
correlations between the electromagnetic fields of QED and the
chromo-electromagnetic fields of QCD in QED+QCD
dynamical-fermion simulations from the QCDSF collaboration.
}

\FullConference{The 36th Annual International Symposium on Lattice Field Theory - LATTICE2018\\
		22-28 July, 2018\\
		Michigan State University, East Lansing, Michigan, USA.}

\begin{document}


\section{Understanding Complexity through Visualisation}
\vspace{-6pt}

Deep insight into the mechanisms giving rise to quantum phenomena can be obtained through the
visualisation of the complex scientific data sets considered in theoretical computational-physics
calculations.  These are insights that would otherwise remain hidden in the typical gigabyte data
sets of modern quantum field theory.  If one can see what's going on, it is easy to share this new
understanding with a broad audience.  In this sense, a picture is worth a thousand equations.

For 20 years, the Centre for the Subatomic Structure of Matter (CSSM) has aimed to reveal the
quantum phenomena of the nontrivial vacuum that underpins every calculation in lattice QCD.  The
focus is on displaying the results of genuine calculations and creating new insights into the
fundamental mechanisms that give rise to the observed phenomena.  Visualisations of these phenomena
not only serve to explain the novel concept of a nontrivial vacuum but also inform and entertain a
diverse audience from researchers in the field to the next generation of science enthusiasts.

These data sets contain a rich diversity of emergent structure and a variety of approaches to
revealing this structure has been explored.  In this presentation, four approaches to
understanding QCD vacuum structure are presented.
%
Each approach is illustrated by animations which are available via YouTube, enabling easy
dissemination through social media.  The original HD-streaming format videos have also been placed
on line.  In the following, references contain active URL's to these animations.

\section{Structure of the QCD Vacuum}
\label{sec:QCDvacuum}
\vspace{-6pt}

Among the earliest of CSSM visualisations are animations of the Euclidean action density and the
topological charge density of gluon field configurations \cite{VisQCDarchive:2002}.  Calculations
were based on pure SU(3) gluon dynamics with the standard Wilson action at $\beta = 6.0$ on a $24^3
\times 36$ lattice.  The first coordinate was used for the time axis creating a $24^2 \times 36$
spatial volume with a lattice spacing of 0.1 fm.  It is these calculations that captured the
attention of Prof Frank Wilczek as he prepared his 2004 Nobel Prize lecture.
Ref.~\cite{NobelAnimation:2004} provides a link to the {\sl QCD Lava Lamp} animation that appeared
in his Nobel Lecture \cite{NobelLecture:2004}.  In support of the Nobel Lecture a web page
incorporating the best algorithms and visualisation techniques of the time was created to engage
the public's interest \cite{VisualQCD:2004}.  Since then these images and animations have appeared
in popular-science publications, YouTube channels \cite{Veritasium:20130430,Veritasium:20130508},
public talks and lectures, conference presentations, journal articles, annual reports, grant
applications, school visits \cite{OriginMass:2008} and newspaper articles.

In this report, a new animation \cite{StructVac:2019} is presented drawing on the legacy
calculations of the Nobel Lecture and providing new insights into the structure of the eight
chromo-electric and eight chromo-magnetic fields.  Correlations between the energy density,
topological charge density and the 16 vector fields are explored.  The HD-streaming
version is available via the link in Ref.~\cite{StructVacHD:2018}.

Calculations of the chromo-electromagnetic fields proceed via the $\mathcal{O}(a^4)$-five-loop
improved field-strength tensor, $F_{\mu\nu}$, designed to ensure the most local $1\times 1$ and
$1\times 2$ clover terms make up 96\% of the loop contributions \cite{BilsonThompson:2002jk}.  To
explore the long-distance nonperturbative phenomena of the vacuum, short-distance fluctuations are
suppressed via 25 sweeps of cooling with the mean-field-improved $\mathcal{O}(a^4)$-three-loop
improved action \cite{BilsonThompson:2003zi}, ensuring the lattice operators are accurate while
preserving topological structure within the gauge fields.  After assigning the electric and
magnetic fields in the Euclidean field-strength tensor for the case with the first coordinate
playing the role of time, the eight fields are extracted via a trace with the Gell-Mann matrices.  

\begin{figure}[t]
\begin{center}
\includegraphics[width=0.45\textwidth]{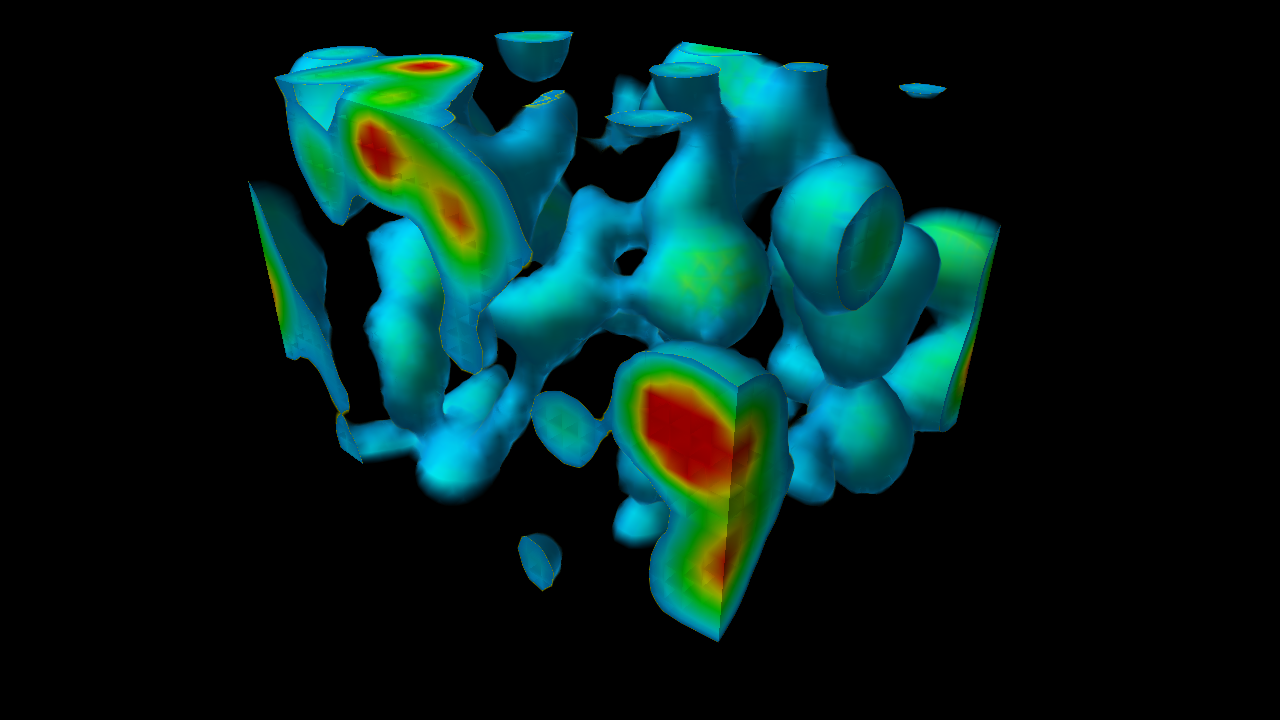}
\quad
\includegraphics[width=0.45\textwidth]{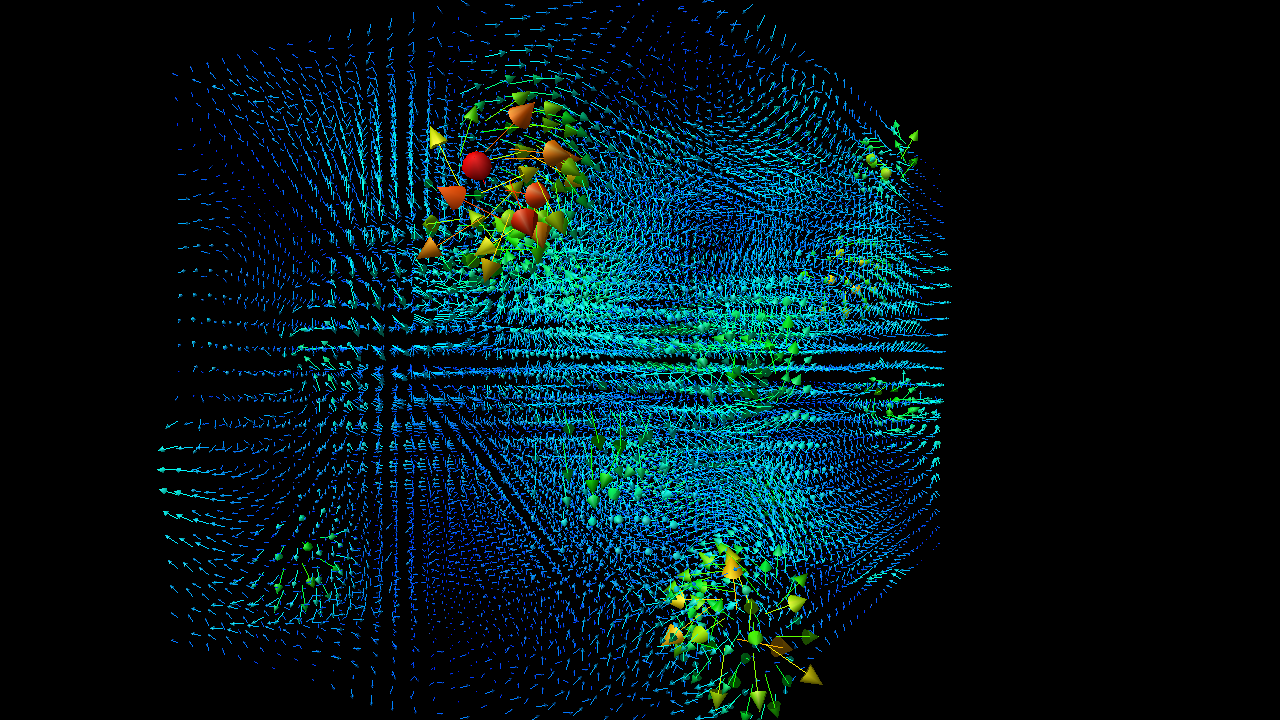}
\vspace{-12pt}
\end{center}
\caption{Frames from the animation of Refs.~\protect\cite{StructVac:2019,StructVacHD:2018}
  illustrating the energy density (left) and one of the chromo-magnetic fields composing the
  nontrivial QCD vacuum.}
\label{fig:StructVac}
\vspace{-6pt}
\end{figure}

The associated animation \cite{StructVac:2019,StructVacHD:2018}, illustrates the reduction of the
energy density of the gluon field in the first 25 frames of the animation and then displays the
time dependence over the 24 frames of the periodic lattice.  Areas of high energy density are
rendered in red and regions of moderate energy density are rendered in blue.  Low-energy regions
are not rendered so we can see into the volume.  Subsequently, the eight chromo-electric and eight
chromo-magnetic gluon fields giving rise to the energy density are
illustrated in Landau gauge \cite{Bonnet:1999mj}.  Here the colour
and length of the arrows describe the magnitude of the vector fields.  Figure~\ref{fig:StructVac}
displays two frames from the animation.  Finally, the animation presents correlations with the
topological charge density with regions of positive density rendered in red through yellow and
regions of negative density rendered blue through cyan.

\vspace{-6pt}
\section{Visualisation of Centre Clusters in Local Polyakov Loops}
\vspace{-6pt}

In Ref.~\cite{Stokes:2013oaa} an anisotropic gauge action was used to explore the evolution of
coherent centre domains in the gluon field under both temperature and the Hybrid Monte Carlo (HMC)
update algorithm.
Centre clusters are defined in terms of the complex-valued local Polyakov loop whose expectation
value acts as an order parameter for the finite temperature phase transition in QCD.  It has an
expectation value of zero in the confined phase and a nonzero expectation value in the deconfined
phase \cite{Gattringer:2010ug}.
The local Polyakov loop is defined as
\[
  L(\vec{x}) := \textrm{Tr}\left(\mathcal{P}\exp \left[i g \int
                     \mathrm{d}x^0 A_0(x) \right]\right) 
              = \textrm{Tr}\prod_{t=1}^{N_t} U_0(t,\vec{x})
              = \rho(\vec{x}) \, e^{i\phi(\vec{x})} \, ,
\]
%
%
%
where it is decomposed into a phase, $\phi(\vec{x})$ and a magnitude,
$\rho(\vec{x})$ in the final expression.
Both the proximity of the phase to one of the cube-roots of one and
the magnitude are considered in visualising the structure of the
centre domains of the gluon field. In either case, the most proximal cube root
of one to the phase is indicated by the use of colour.

\begin{figure}[t]
\begin{center}
  \subfloat[Defined by phase]{
    \includegraphics[width=0.3\textwidth]{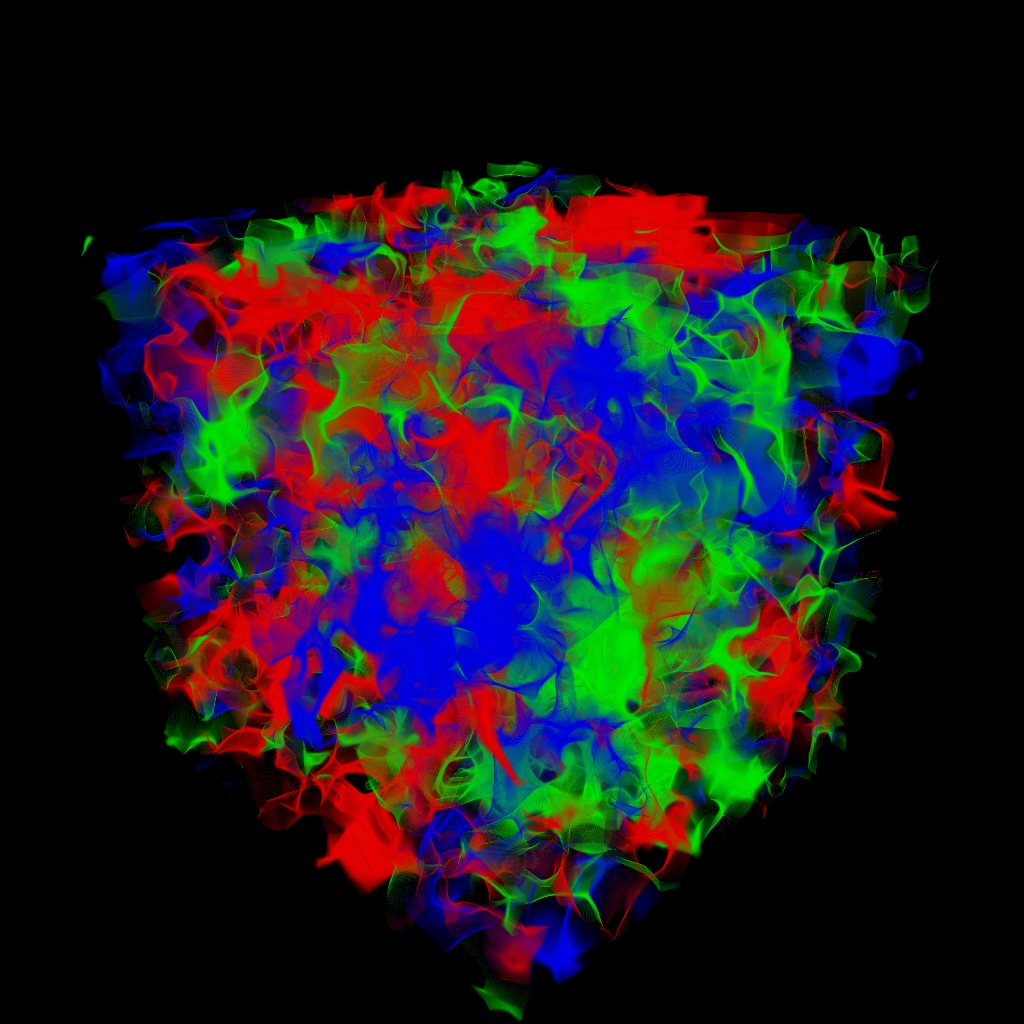}
    \label{fig:evolutionsmear1}
  }
  \qquad\qquad
  \subfloat[Defined by magnitude]{
    \includegraphics[width=0.3\textwidth]{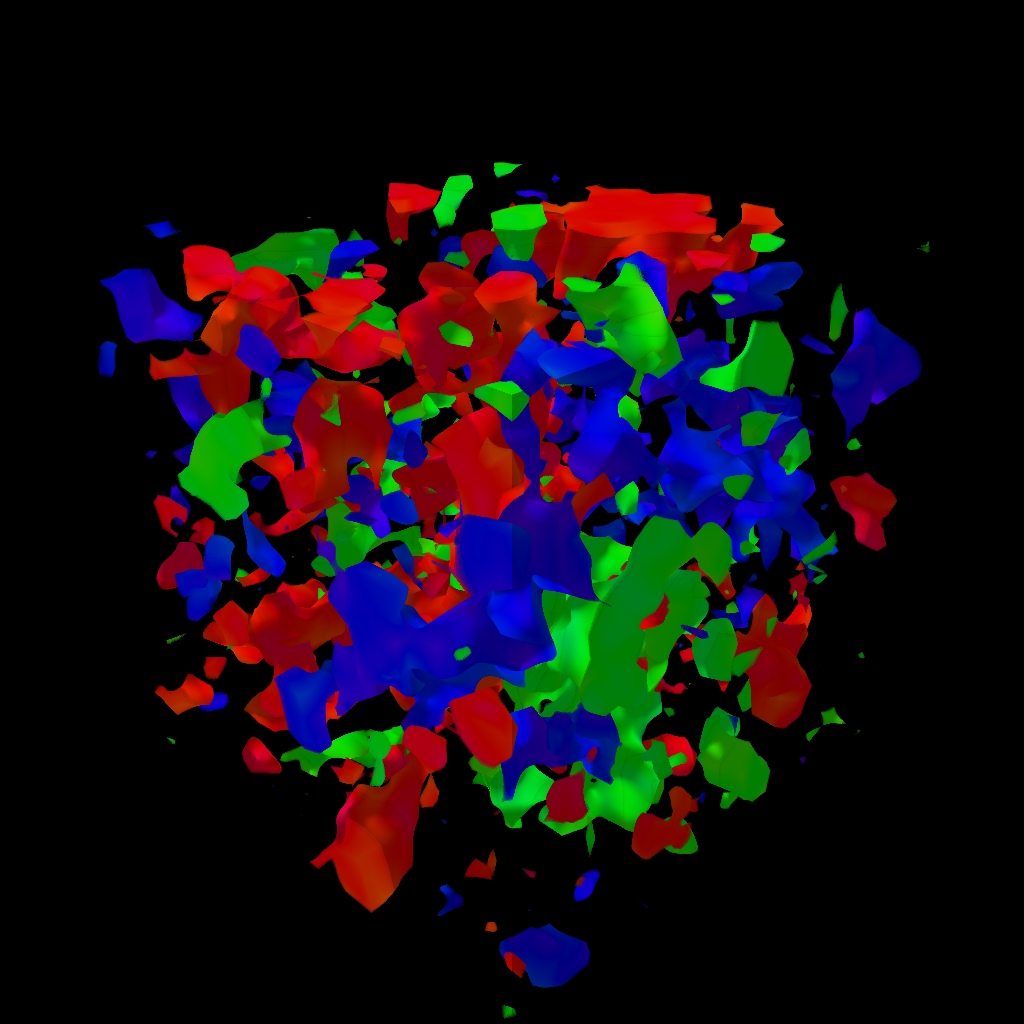}
    \label{fig:evolutionsmear2}
  }
\end{center}
\vspace{-12pt}
  \caption{Centre clusters on a gauge field configuration at $T = 0.89(1) \, T_C$. Similar
    structure is revealed for centre clusters defined as regions proximal to a single centre phase,
    and as regions with a larger magnitude. Four sweeps of stout-link smearing are applied to the
    gauge links prior to calculating the Polyakov loops.  The length of each side of the cubic
    volume is 2.4 fm. }
  \label{fig:evolutionsmear}
\vspace{-6pt}
\end{figure}

To investigate the larger-scale behaviour of the clusters, small scale noise is
removed from the visualisation by performing four sweeps of stout-link smearing
prior to calculating the Polyakov loops.  Fig.~\ref{fig:evolutionsmear} illustrates
two different ways of defining the clusters. In Fig.~\ref{fig:evolutionsmear2},
clusters are rendered where the phase $\phi(\vec x)$ is within some small window
around each centre phase, and the rest of the volume is rendered transparent.
The evolution of these clusters with HMC simulation time is presented in
Refs.~\cite{CentreDomainsPhase:2014,CentreDomainsPhaseHD:2014}, showing how centre
clusters are slowly moving. The animation \cite{CentreDomainsPhase:2014,CentreDomainsPhaseHD:2014}
reveals correlations in the centre clusters persisting for approximately 5 seconds corresponding to
25 HMC trajectories.  The temperature dependence of the centre-cluster structure is also explored
in these animations where a single phase eventually dominates above the critical temperature.
In Fig.~\ref{fig:evolutionsmear2}, and the animation of Ref.~\cite{CentreDomainsMag:2014,CentreDomainsMagHD:2014},
the clusters are rendered where the magnitude $\rho(\vec x)$ is above some
threshold.

These fundamental structures support the concept of hadrons being described as a quark core dressed
by a meson cloud.  Within these coherent centre domains, colour-singlet quark-antiquark pairs or
three-quark triplets have a finite energy and are spatially correlated.  Thus, these fundamental
domains govern the size of the quark cores of hadrons.  As one domain dominates the vacuum above
the critical temperature, the correlation length diverges and quarks become deconfined.

\vspace{-6pt}
\section{Structure of Projected Centre Vortices in the Nontrivial QCD Vacuum}
\vspace{-6pt}

Recent research is now exposing the centre-vortex structure of nonperturbative gluon-field
configurations to be the most fundamental aspect of nonperturbative vacuum structure giving rise to
both confinement and dynamical chiral-symmetry breaking.
Removal of SU(3) centre vortices removes confinement, while consideration of the vortices alone
provides confinement \cite{Langfeld:2003ev}.  Similarly, removal of vortices suppresses the
infrared enhancement of the gluon propagator, while consideration of the vortices alone provides
the well-known infrared enhancement \cite{Biddle:2018dtc}.  Studies of the nonperturbative quark
propagator provide evidence that centre vortices underpin dynamical chiral symmetry breaking
\cite{Trewartha:2015nna}, and the removal of centre vortices from the gluon fields restores chiral
symmetry \cite{Trewartha:2017ive}.  Centre vortices are the seeds of dynamical chiral symmetry
breaking.

In light of the importance of these most fundamental aspects of QCD vacuum structure,
visualisations of the complex structures formed by the projected centre vortices in SU(3) gauge
theory have been presented in Ref.~\cite{VortexStructure}.  Here we explore their correlation with
the topological charge density of the gluon fields and present a new animation
\cite{Vortices:2019,VorticesHD:2018} showcasing a prevalence of branching points in the flow of
centre charge that look like monopole or anti-monopole contributions.  Details of the
identification of projected centre vortices and the techniques used to render them are described in
Ref.~\cite{VortexStructure} of these proceedings.  The topological charge  is determined using the
techniques of Sec.~\ref{sec:QCDvacuum} with eight sweeps of $\mathcal{O}(a^4)$-three-loop
improved cooling \cite{BilsonThompson:2003zi}.

\begin{figure}[t]
\begin{center}
\includemedia[
        noplaybutton,
	3Dtoolbar,
	3Dmenu,
	3Dviews=PlaqLinkTopQ_CFG95_T02.vws,
	3Dcoo  = 10 10 20, 
	3Dc2c  = 0 1 0,    
	3Droo  = 70,       
	3Droll = 270,      
	3Dlights=CAD,
	width=\textwidth, 
]{\includegraphics{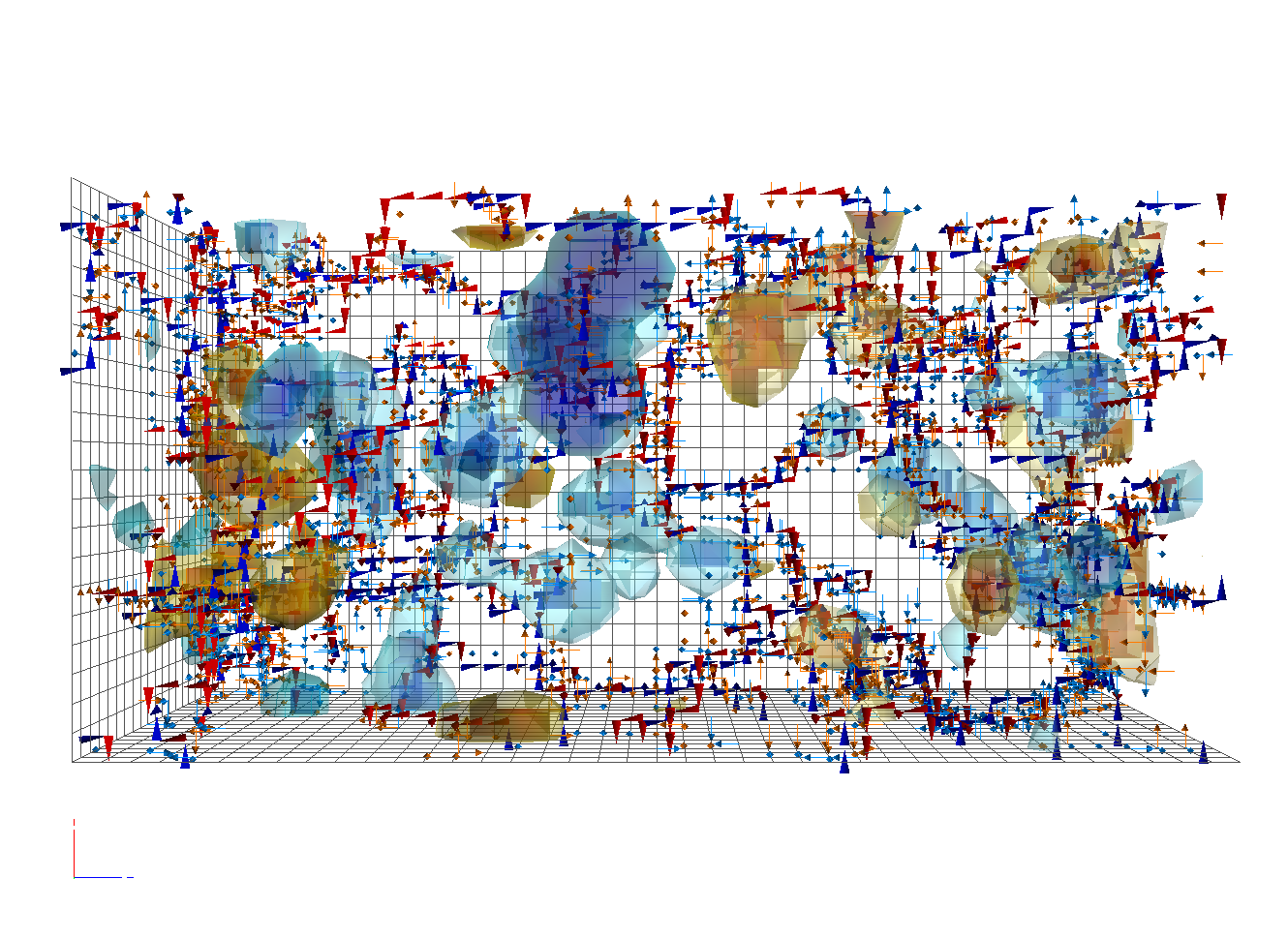}}{PlaqLinkTopQ_CFG95_T02.u3d}
\end{center}
\vspace{-24pt}
\caption{ {\bf The projected centre-vortex structure of the gluon field.}  The flow of centre
  charge is illustrated by the red and blue jets.  The motion of the jets through time-oriented
  plaquettes is indicated by the orange and cyan arrows.  The correlation of the vortex lines with
  the topological charge density is examined by volume-rendering the topological charge density in
  a translucent manner.  Additional details are available in Ref.~\cite{VortexStructure}.  (Click
  on the image to activate it in Adobe Reader. Click and drag to rotate, Ctrl-click to translate,
  Shift-click or mouse wheel to zoom. Right click to access the ``Views'' menu.)  }
\vspace{-6pt}
\label{fig:PlaqLinkTopQ_CFG95_T02.u3d}
\end{figure}

The structure of projected centre vortices and their correlation with the topological charge
density is illustrated\footnote{To interact with Fig.~\ref{fig:PlaqLinkTopQ_CFG95_T02.u3d}, open
  this pdf document in Adobe Reader 9 or later.  Linux users should install
  \href{ftp://ftp.adobe.com/pub/adobe/reader/unix/9.x/9.4.1/enu/}{Adobe acroread version 9.4.1},
  the last edition to have full 3D support.  From the ``Edit'' menu, select ``Preferences...'' and
  ensure ``3D \& Multimedia'' is enabled and ``Enable double-sided rendering'' is selected.} in
Fig.~\ref{fig:PlaqLinkTopQ_CFG95_T02.u3d}.  Inspection of the vortices reveals the flow of centre
charge, intersection points and a prevalence of branching points that look like monopole or
anti-monopole contributions.  An engaging presentation suitable for outreach activities is
presented in the animation of Ref.~\cite{Vortices:2019} with the original HD-streaming animation
available via Ref.~\cite{VorticesHD:2018}.

\vspace{-6pt}
\section{Interplay of QED and QCD in the Nontrivial Vacuum}
\vspace{-6pt}

The QCDSF collaboration have generated gauge field configurations incorporating the dynamical
effects of both QCD and QED \cite{Horsley:2015eaa,Horsley:2015vla}.  As the dynamics of the QCD
fields include the creation of quark-antiquark pairs, their electric charge -- enhanced by a factor
of 10 in the simulations -- participates in a dynamical manner through QED interactions.  Here we
consider the $24^3 \times 48$ lattice at the flavor-symmetric point with a lattice spacing of 0.068
fm \cite{Horsley:2015eaa}.  The calculations of the fields proceed in a manner analogous to that
presented in Sec.~\ref{sec:QCDvacuum}.

\begin{figure}[t]
\begin{center}
\includegraphics[width=0.45\textwidth]{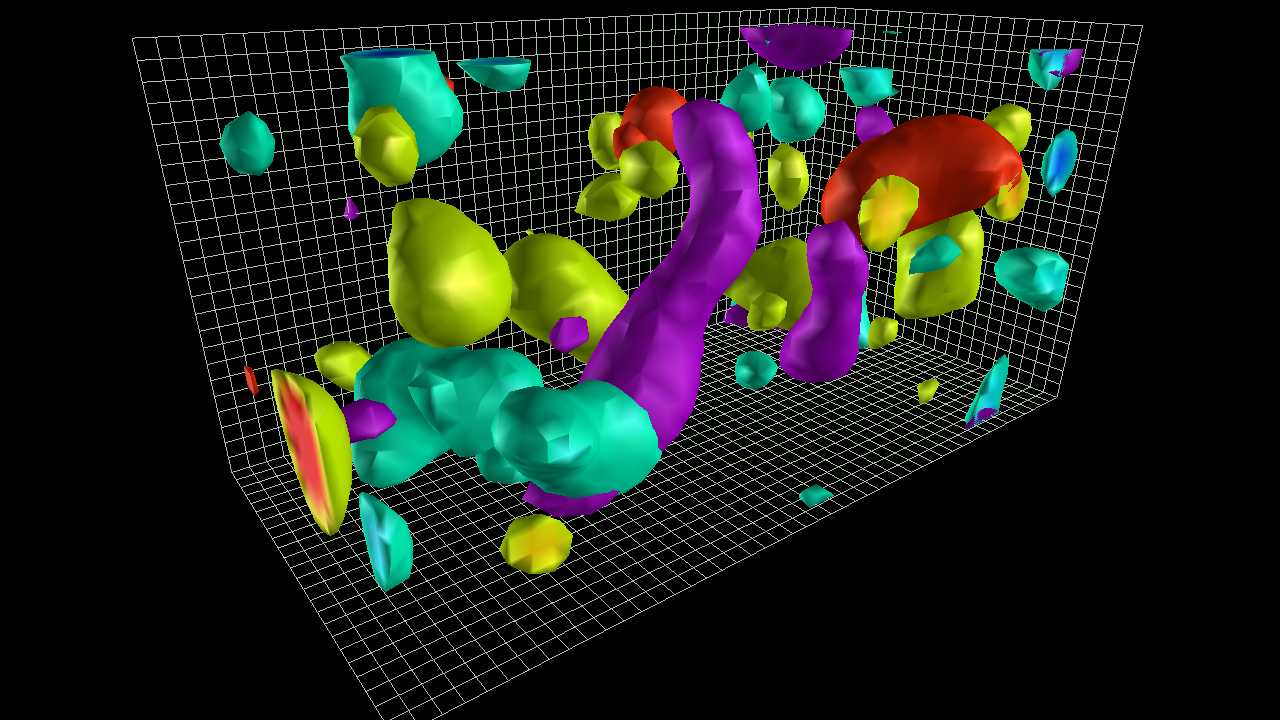}
\quad
\includegraphics[width=0.45\textwidth]{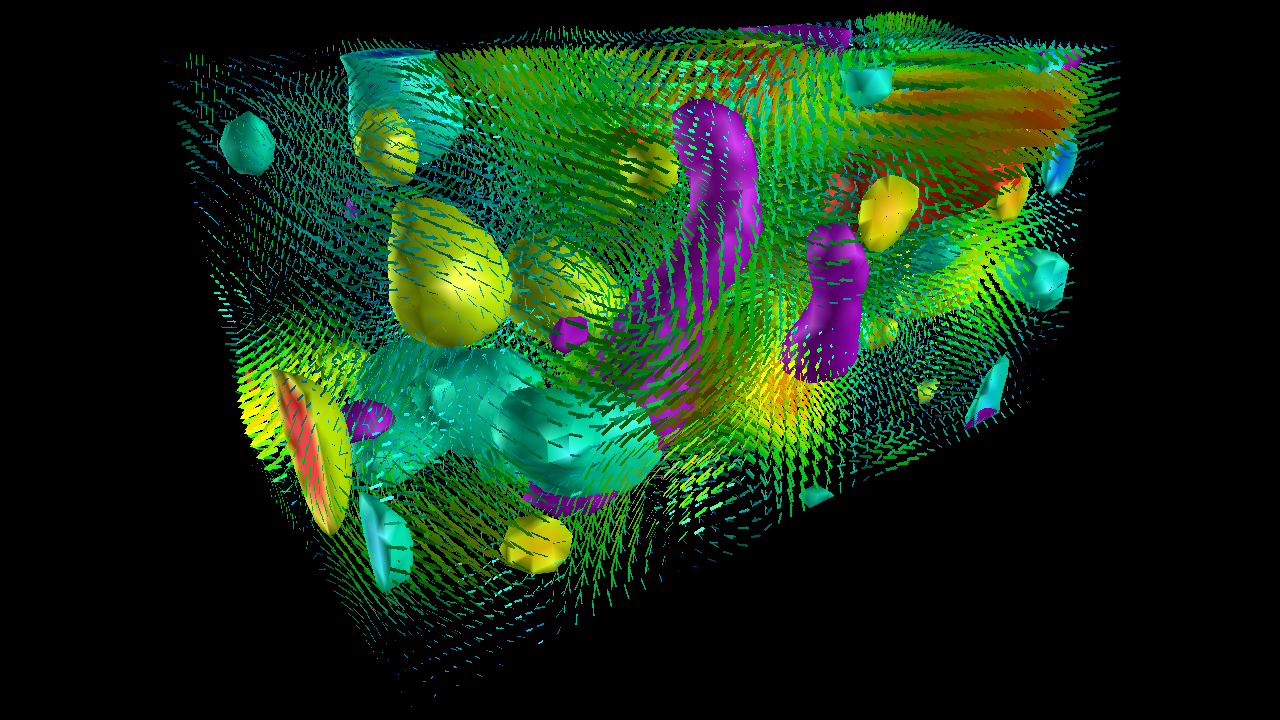}
\vspace{-12pt}
\end{center}
\caption{Frames from the animation of Refs.~\protect\cite{QEDnQCD:2019,QEDnQCDHD:2018} illustrating
  the topological charge density of QCD and the electric charge density of QED (left) with the QED
  magnetic field superposed (right).  Regions of significant topological charge density are
  rendered in red through yellow (blue through cyan) for positive (negative) values of the
  density. Regions of large electric charge density are rendered in red (purple) for positive
  (negative) charge and the associated magnetic field is illustrated by the vector field.  }
\label{fig:QEDnQCD}
\end{figure}

There is an interesting relationship between the electric charge density, obtained through the
divergence of the electric field, and the topological charge density of QCD.  As the electric
charge moves, magnetic fields are created in the vacuum.  The movement of electric charge relative
to the magnetic field, particularly when that charge is influenced by the topological charge
density of QCD is of particular interest.
Figure \ref{fig:QEDnQCD} presents two frames from the animation of Refs.~\cite{QEDnQCD:2019,QEDnQCDHD:2018}.
In this animation the topological charge density is compared with the electric charge density of
quark-antiquark pairs and the magnetic field they create.

\begin{thebibliography}{10}

\bibitem{VisQCDarchive:2002}
D.~B. Leinweber, \emph{Visual QCD Archive}, 2002.
\newblock
  \href{http://www.physics.adelaide.edu.au/theory/staff/leinweber/VisualQCD/QCDvacuum/}{\small
  http://www.physics.adelaide.edu.au/theory/staff/leinweber/VisualQCD/QCDvacuum/}.

\bibitem{NobelAnimation:2004}
D.~Leinweber, \emph{QCD Lava Lamp}, 2004.
\newblock
  \href{http://www.physics.adelaide.edu.au/theory/staff/leinweber/VisualQCD/QCDvacuum/su3b600s24t36cool30action.gif}{\small
  http://www.physics.adelaide.edu.au/theory/staff/leinweber/VisualQCD/QCDvacuum/
  su3b600s24t36cool30action.gif}.

\bibitem{NobelLecture:2004}
F.~Wilczek, \emph{2004 Nobel Prize Lecture}, 2004.
\newblock
  \href{https://www.nobelprize.org/prizes/physics/2004/wilczek/lecture/}{\small
  https://www.nobelprize.org/prizes/physics/2004/wilczek/lecture/}.

\bibitem{VisualQCD:2004}
D.~B. Leinweber, \emph{Visualizations of Quantum Chromodynamics}, 2004.
\newblock
  \href{http://www.physics.adelaide.edu.au/theory/staff/leinweber/VisualQCD/Nobel/}{\small
  http://www.physics.adelaide.edu.au/theory/staff/leinweber/VisualQCD/Nobel/}.

\bibitem{Veritasium:20130430}
D.~Muller, \emph{Empty Space is NOT Empty}, 2013.
\newblock \href{https://www.youtube.com/watch?v=S1tFT4smd6E}{\small
  https://www.youtube.com/watch?v=S1tFT4smd6E}.

\bibitem{Veritasium:20130508}
D.~Muller, \emph{Your Mass is NOT From the Higgs Boson}, 2013.
\newblock \href{https://www.youtube.com/watch?v=Ztc6QPNUqls}{\small
  https://www.youtube.com/watch?v=Ztc6QPNUqls}.

\bibitem{OriginMass:2008}
D.~B. Leinweber, \emph{Origin of Mass}, 2008.
\newblock
  \href{http://www.physics.adelaide.edu.au/theory/staff/leinweber/VisualQCD/OriginMass/}{\small
  http://www.physics.adelaide.edu.au/theory/staff/leinweber/VisualQCD/OriginMass/}.

\bibitem{StructVac:2019}
D.~B. Leinweber and E.~Puckridge, \emph{Structure of the QCD Vacuum}, 2019.
\newblock \href{https://youtu.be/WZgZI5vymiM}{\small
  https://youtu.be/WZgZI5vymiM}.

\bibitem{StructVacHD:2018}
D.~B. Leinweber and E.~Puckridge, \emph{Structure of the QCD Vacuum}, 2018.
\newblock
  \href{http://www.physics.adelaide.edu.au/cssm/lattice/QCDvacuum/ChromoElectroMagneticFields.mp4}{\small
  http://www.physics.adelaide.edu.au/cssm/lattice/QCDvacuum/ChromoElectroMagneticFields.mp4}.

\bibitem{BilsonThompson:2002jk}
S.~O. Bilson-Thompson, D.~B. Leinweber and A.~G. Williams, \emph{{Highly
  improved lattice field strength tensor}},
  \href{https://doi.org/10.1016/S0003-4916(03)00009-5}{\emph{Annals Phys.}
  {\bfseries 304} (2003) 1}
  [\href{https://arxiv.org/abs/hep-lat/0203008}{{\ttfamily hep-lat/0203008}}].

\bibitem{BilsonThompson:2003zi}
S.~O. Bilson-Thompson et~al., \emph{{Comparison of |Q| = 1 and |Q| = 2
  gauge-field configurations on the lattice four-torus}},
  \href{https://doi.org/10.1016/j.aop.2003.12.011}{\emph{Annals Phys.}
  {\bfseries 311} (2004) 267}
  [\href{https://arxiv.org/abs/hep-lat/0306010}{{\ttfamily hep-lat/0306010}}].

\bibitem{Bonnet:1999mj}
F.~D.~R. Bonnet, P.~O. Bowman, D.~B. Leinweber, A.~G. Williams and D.~G.
  Richards, \emph{{Discretization errors in Landau gauge on the lattice}},
  \href{https://doi.org/10.1071/PH99047}{\emph{Austral. J. Phys.} {\bfseries
  52} (1999) 939} [\href{https://arxiv.org/abs/hep-lat/9905006}{{\ttfamily
  hep-lat/9905006}}].

\bibitem{Stokes:2013oaa}
F.~M. Stokes, W.~Kamleh and D.~B. Leinweber, \emph{{Visualizations of coherent
  center domains in local Polyakov loops}},
  \href{https://doi.org/10.1016/j.aop.2014.05.002}{\emph{Annals Phys.}
  {\bfseries 348} (2014) 341}
  [\href{https://arxiv.org/abs/1312.0991}{{\ttfamily 1312.0991}}].

\bibitem{Gattringer:2010ug}
C.~Gattringer and A.~Schmidt, \emph{{Center clusters in the Yang-Mills
  vacuum}}, \href{https://doi.org/10.1007/JHEP01(2011)051}{\emph{JHEP}
  {\bfseries 01} (2011) 051} [\href{https://arxiv.org/abs/1011.2329}{{\ttfamily
  1011.2329}}].

\bibitem{CentreDomainsPhase:2014}
F.~M. Stokes, W.~Kamleh and D.~B. Leinweber, \emph{Centre Domains in the QCD
  Vacuum - Smeared Phase}, 2014.
\newblock \href{https://youtu.be/KkiOQOOb69k}{\small
  https://youtu.be/KkiOQOOb69k}.

\bibitem{CentreDomainsPhaseHD:2014}
F.~M. Stokes, W.~Kamleh and D.~B. Leinweber, \emph{Centre Domains in the QCD
  Vacuum - Smeared Phase}, 2014.
\newblock
  \href{http://www.physics.adelaide.edu.au/cssm/lattice/centreclusters/smeared-phase.mp4}{\small
  http://www.physics.adelaide.edu.au/cssm/lattice/centreclusters/smeared-phase.mp4}.

\bibitem{CentreDomainsMag:2014}
F.~M. Stokes, W.~Kamleh and D.~B. Leinweber, \emph{Centre Domains in the QCD
  Vacuum - Smeared Magnitude}, 2014.
\newblock \href{https://youtu.be/T4sRON6uOz0}{\small
  https://youtu.be/T4sRON6uOz0}.

\bibitem{CentreDomainsMagHD:2014}
F.~M. Stokes, W.~Kamleh and D.~B. Leinweber, \emph{Centre Domains in the QCD
  Vacuum - Smeared Magnitude}, 2014.
\newblock
  \href{http://www.physics.adelaide.edu.au/cssm/lattice/centreclusters/smeared-magnitude.mp4}{\small
  http://www.physics.adelaide.edu.au/cssm/lattice/centreclusters/smeared-magnitude.mp4}.

\bibitem{Langfeld:2003ev}
K.~Langfeld, \emph{{Vortex structures in pure SU(3) lattice gauge theory}},
  \href{https://doi.org/10.1103/PhysRevD.69.014503}{\emph{Phys. Rev.}
  {\bfseries D69} (2004) 014503}
  [\href{https://arxiv.org/abs/hep-lat/0307030}{{\ttfamily hep-lat/0307030}}].

\bibitem{Biddle:2018dtc}
J.~C. Biddle, W.~Kamleh and D.~B. Leinweber, \emph{{Gluon propagator on a
  center-vortex background}},
  \href{https://doi.org/10.1103/PhysRevD.98.094504}{\emph{Phys. Rev.}
  {\bfseries D98} (2018) 094504}
  [\href{https://arxiv.org/abs/1806.04305}{{\ttfamily 1806.04305}}].

\bibitem{Trewartha:2015nna}
D.~Trewartha, W.~Kamleh and D.~Leinweber, \emph{{Evidence that centre vortices
  underpin dynamical chiral symmetry breaking in SU(3) gauge theory}},
  \href{https://doi.org/10.1016/j.physletb.2015.06.025}{\emph{Phys. Lett.}
  {\bfseries B747} (2015) 373}
  [\href{https://arxiv.org/abs/1502.06753}{{\ttfamily 1502.06753}}].

\bibitem{Trewartha:2017ive}
D.~Trewartha, W.~Kamleh and D.~Leinweber, \emph{{Centre vortex removal restores
  chiral symmetry}}, \href{https://doi.org/10.1088/1361-6471/aa9443}{\emph{J.
  Phys.} {\bfseries G44} (2017) 125002}
  [\href{https://arxiv.org/abs/1708.06789}{{\ttfamily 1708.06789}}].

\bibitem{VortexStructure}
J.~C. Biddle, W.~Kamleh and D.~B. Leinweber, \emph{{Visualizations of Centre
  Vortex Structure in Lattice Simulations}}, {\emph{PoS} {\bfseries
  LATTICE2018} (2018) 256} [\href{https://arxiv.org/abs/1903.07767}{{\ttfamily
  1903.07767}}].

\bibitem{Vortices:2019}
J.~Biddle, W.~Kamleh, D.~Leinweber and H.~Piercy, \emph{Centre Vortices in the
  Gluon Field of the QCD Vacuum}, 2019.
\newblock \href{https://youtu.be/CDdmx989quA}{\small
  https://youtu.be/CDdmx989quA}.

\bibitem{VorticesHD:2018}
J.~Biddle, W.~Kamleh, D.~Leinweber and H.~Piercy, \emph{Centre Vortices in the
  Gluon Field of the QCD Vacuum}, 2018.
\newblock
  \href{http://www.physics.adelaide.edu.au/cssm/lattice/QCDvacuum/CentreVorticesTopQ.mp4}{\small
  http://www.physics.adelaide.edu.au/cssm/lattice/QCDvacuum/CentreVorticesTopQ.mp4}.

\bibitem{Horsley:2015eaa}
R.~Horsley et~al., \emph{{Isospin splittings of meson and baryon masses from
  three-flavor lattice QCD + QED}},
  \href{https://doi.org/10.1088/0954-3899/43/10/10LT02}{\emph{J. Phys.}
  {\bfseries G43} (2016) 10LT02}
  [\href{https://arxiv.org/abs/1508.06401}{{\ttfamily 1508.06401}}].

\bibitem{Horsley:2015vla}
R.~Horsley et~al., \emph{{QED effects in the pseudoscalar meson sector}},
  \href{https://doi.org/10.1007/JHEP04(2016)093}{\emph{JHEP} {\bfseries 04}
  (2016) 093} [\href{https://arxiv.org/abs/1509.00799}{{\ttfamily
  1509.00799}}].

\bibitem{QEDnQCD:2019}
J.~Charvetto, W.~Kamleh, D.~Leinweber, R.~Young and J.~Zanotti, \emph{Interplay
  of Quantum Electrodynamics and Quantum Chromodynamics in the Nontrivial
  Vacuum}, 2019.
\newblock \href{https://youtu.be/9TJe1Pr5c9Q}{\small
  https://youtu.be/9TJe1Pr5c9Q}.

\bibitem{QEDnQCDHD:2018}
J.~Charvetto, W.~Kamleh, D.~Leinweber, R.~Young and J.~Zanotti, \emph{Interplay
  of Quantum Electrodynamics and Quantum Chromodynamics in the Nontrivial
  Vacuum}, 2018.
\newblock
  \href{http://www.physics.adelaide.edu.au/cssm/lattice/QEDnQCD/QEDnQCDvacuum.mp4}{\small
  http://www.physics.adelaide.edu.au/cssm/lattice/QEDnQCD/QEDnQCDvacuum.mp4}.

\end{thebibliography}

\providecommand{\href}[2]{#2}\begingroup\raggedright\endgroup

\end{document}